\documentstyle[pra,aps,epsf,floats,graphicx]{revtex}
\def\sb#1{$_{#1}$}
\def\sp#1{$^{#1}$}
\def\ia{~\AA$^{-1}$}
\def\a2{~\AA$^{2}$}
\begin{document}

\draft
\wideabs{

\title{Local structure of In\sb{0.5}Ga\sb{0.5}As from joint high-resolution and 
differential pair distribution function analysis
}

\author{V. Petkov, I-K. Jeong, F. Mohiuddin-Jacobs, Th. Proffen, S.J.L. Billinge}
\address{Department of Physics and Astronomy and Center for Fundamental Materials Research, Michigan State \\
University, East Lansing, MI-48823-1116}
\author{W. Dmowski}
\address{Laboratory for Research on the Structure of Matter and Department of Materials Science and Engineering, 
University of Pennsylvania, 9201 Walnut Street, Philadelphia, PA 19104}

\date{17 November 1999}
\maketitle

\begin{abstract}
High resolution total and indium differential atomic pair distribution functions (PDFs) for 
In\sb{0.5}Ga\sb{0.5}As alloys have been obtained by high energy and anomalous x-ray diffraction 
experiments, respectively.  The first peak in the total PDF is resolved as a doublet  due to 
the presence of two distinct bond lengths, In-As and Ga-As. The In differential PDF, 
which  involves only atomic pairs containing In,  yields chemical specific information 
and helps ease the structure data interpretation. Both PDFs have been fit with structure 
models and the way in that  the underlying cubic zinc-blende lattice of In\sb{0.5}Ga\sb{0.5}As
semiconductor alloy distorts locally to  accommodate the distinct In-As and Ga-As bond 
lengths present has been quantified.  
\end{abstract}


}

\section{Introduction}

Ternary semiconductor alloys, such as In\sb{x}Ga\sb{1-x}As, are 
technologically important because they allow the 
semiconductor band-gap to be varied continuously 
between the band-gap values of the end members, GaAs 
and InAs, by varying the composition, $x$ \cite{[1],[2]}. This has 
made the technological characteristics, physical 
properties and structure  of In\sb{x}Ga\sb{1-x}As alloys a subject of 
numerous experimental and theoretical investigations.  It 
has been found that the lattice parameters of the alloys 
well interpolate between those of the end members 
which is consistent with  the so-called Vegard's law.  
According to Vegard's law, the  structure of alloys   
adjusts itself so that the individual bond lengths are 
taking equal, compositionally averaged, values and the 
bond angles remain unperturbed from their ideal values 
for any alloy composition. For this reason, all structure 
dependent properties of In\sb{x}Ga\sb{1-x}As alloys, including 
electronic band-structure, are often calculated within the 
virtual crystal approximation (VCA).  In this 
approximation the alloy is assumed to be a perfect 
crystal with all atoms sitting on ideal lattice sites and site 
occupancies given by the {\it average} alloy composition. 
Both GaAs and InAs have the zinc-blende structure 
($F\overline{4} 3m$) where In and Ga atoms occupy two 
interpenetrating face-centered-cubic (fcc) lattices and are 
tetrahedrally coordinated to each other \cite{[3]}. Accordingly, 
the VCA assumes that In\sb{x}Ga\sb{1-x}As alloys have the same 
zinc-blende structure and, furthermore, the first 
neighbor interatomic distances (i.e. In-As and Ga-As 
bonds lengths), bond ionicity, atomic potential etc. take 
some {\it average} values for any composition $x$. However, 
both extended x-ray absorption fine structure (XAFS) 
experiments \cite{[4]} and theory \cite{[5]} have shown that Ga-As 
and In-As bonds do not take some {\it average} value but 
remain close to their {\it natural} lengths L\sp{o}\sb{Ga-As} = 2.437~\AA\ 
and L\sp{o}\sb{In-As} =  2.610~\AA\ in the alloy. This behavior is close 
to the so-called Pauling model \cite{[6]} which assumes that 
the bond length between a given atomic pair in an alloy 
is more or less a constant, independent on composition $x$.  
This finding shows that the zinc-blende lattice of In\sb{x}Ga\sb{1-x}As
is significantly deformed to 
accommodate the two-distinct Ga-As and In-As bond 
lengths present. Also, the deformation seems to be 
confined {\it locally} since the {\it average} crystal symmetry and 
structure is still of the cubic zinc-blende type as 
manifested by the Bragg diffraction patterns. 

It is well recognized that {\it local} structural 
distortions and associated fluctuations in atomic 
potentials significantly affect the properties of materials 
and, therefore, should be accounted for in more realistic 
theoretical calculations \cite{[7]}. Thus it seems there  is a 
clear need for more detailed determination of the local 
structure of In\sb{x}Ga\sb{1-x}As semiconductor alloys which then 
may be used as an improved quality structure input to 
theoretical calculations. A number of authors 
\cite{[4],[5],[8],[9],silve;epl95} have already proposed model structures for 
these alloys but there has been limited experimental 
evidence to date.  This prompted us to undertake an 
extensive experimental study of the local atomic 
structure of In-Ga-As semiconductor alloys. 

The technique of choice for studying the local 
structure of semiconductor alloys has been XAFS  
\cite{[4],[10],[11],[12]}. However, XAFS provides information about 
the immediate atomic ordering (first and sometimes 
second coordination shells) and all longer-ranged 
structural features remain hidden. To remedy this 
shortcoming we have taken the alternative approach of 
obtaining atomic pair distribution functions from x-ray 
diffraction data.

The atomic pair distribution function (PDF) is 
the instantaneous atomic number density-density 
correlation function which describes the atomic 
arrangement in materials \cite{[13]}.  It is the sine Fourier 
transform of the experimentally observable total 
structure factor obtained from a powder diffraction 
experiment. Since the total structure factor, as defined 
in Ref. 14, includes both the Bragg scattered intensities 
and the diffuse scattering part of the diffraction spectrum 
its Fourier associate, the PDF, yields both the {\it local} and 
{\it average} atomic structure of materials. By contrast an 
analysis of the Bragg scattered intensities alone, by a 
Rietveld type analysis \cite{[15]} for instance, yields the 
{\it average} crystal structure only. Determining the PDF has 
been the approach of choice for characterizing glasses, 
liquids and amorphous materials for a long time \cite{[16]}. 
However, its wide spread application to crystalline 
materials, where some deviation from the {\it average} 
structure is expected to take place, has been relatively 
recent \cite{[17]}. The present study is a further step along this 
line. 

Very high real space resolution is required to 
differentiate the distinct Ga-As and In-As bond lengths 
present in In\sb{x}Ga\sb{1-x}As alloys. High resolution is attained 
by obtaining the total structure factor S(Q), where 
$Q=4\pi \sin\theta/\lambda$ is the magnitude of the wave vector, to 
very a high value of $Q$ ( $Q > 40$~\AA\sp{-1}). Here, $2\theta$ is the 
angle between the directions of the incoming and 
outgoing radiation beams and $\lambda$ is the wavelength of the 
radiation used. Recently, we carried out a high energy 
(60 keV; $\lambda=0.206$~\AA ) x-ray diffraction experiment and 
succeeded in obtaining PDFs for In\sb{x}Ga\sb{1-x}As crystalline 
materials  ($x=0$, 0.13, 0.33, 0.5, 0.83, 1) of resolution high 
enough to differentiate Ga-As and In-As first atomic 
neighbor distances present \cite{[18]}. An analysis of the 
experimental data (see Fig. 4 in Ref. \cite{[18]}) showed 
that the local disorder in In\sb{x}Ga\sb{1-x}As materials peaks at a 
composition $x=0.5$. This observation suggested In\sb{0.5}Ga\sb{0.5} 
  as the most appropriate candidate for studying the 
effect of bond-length mismatch on the local structure  of 
In-Ga-As family of semiconductor alloys.  An important 
detail of the high energy experiments carried out is that 
low temperature (10 K) was used to minimize the 
thermal vibration in the samples, and hence increase the 
sensitivity to intrinsic atomic displacements. This left 
open the question about the impact of temperature on the 
local structure of In\sb{x}Ga\sb{1-x}As alloys and necessitated the 
carrying out of an complimentary experiment at 
temperatures considerably higher than 10 K. To partially 
compensate for the inevitable loss of resolution from  the 
thermal broadening of atomic pair we carried out an 
anomalous scattering experiment and determined the In 
differential PDF at room temperature. 

In the present paper we report the high energy 
low temperature and anomalous scattering (In edge) 
room temperature experiments on the In\sb{0.5}Ga\sb{0.5}As.
The experimental total and In differential PDFs have 
been fit with  structure models and the way in 
which the zinc-blende lattice {\it locally} distorts to 
accommodate the two distinct Ga-As and In-As bonds 
present has been quantified. 

\section{EXPERIMENTAL DETAILS}

\subsection{Sample preparation}

The In\sb{0.5}Ga\sb{0.5}As alloy was prepared by mixing 
reagent grade GaAs and InAs powders in the proper amounts.
These were sealed under vacuum in quartz tubes.  The powders were
heated above the liquidus and held for 3 hours to melt them, 
followed by quenching into cold water.
The resulting inhomogeneous alloys were ground, resealed in
quartz tubes under vacuum, and annealed at a temperature just below the
solidus of the alloy for 72-96 hours. 
This procedure was repeated until the samples were homogeneous
as determined from an x-ray diffraction measurement.
The sample for high-energy x-ray 
diffraction measurements was a thin layer of the powder 
held between Kapton foils. The thickness of the layer 
was optimized to achieve a sample absorption $\mu$t $\sim 1$ for 
the 60~KeV x-rays. A standard 
sample holder with a cavity of  rectangular shape (2 cm 
x 4 cm) and depth of 0.5 mm was used with the 
anomalous x-ray diffraction experiments. The powder 
was loaded into the cavity to avoid any texture 
formation and its extended surface left openly exposed to 
the x-ray beam.         

\subsection{High-energy x-ray diffraction experiments}

\begin{figure}[!tb]
    \centering
  \includegraphics[angle=0,width=3.0in]{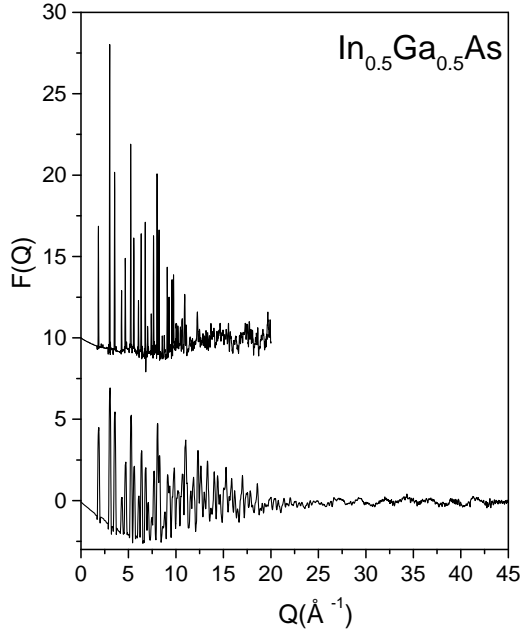}
  \caption{
Total (lower part) and In differential
               (upper part, offset  for clarity) reduced 
               structure factors for In\sb{0.5}Ga\sb{0.5}.  
}
\label{1}
\end{figure}

\begin{figure}[!tb]
    \centering
  \includegraphics[angle=0,width=3.0in]{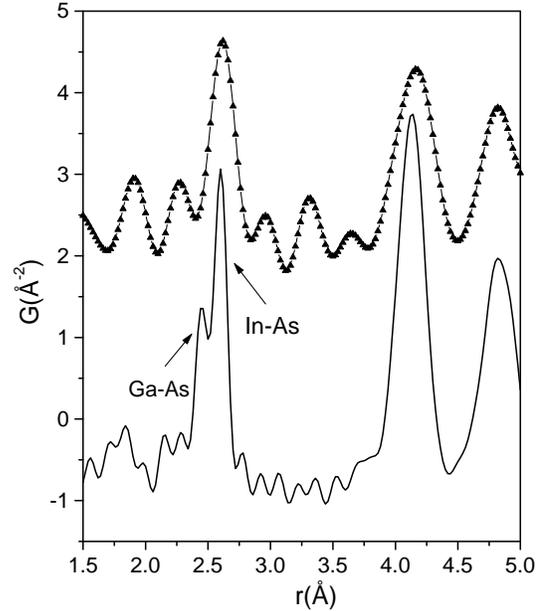}
\caption{The reduced  total (full line) and In 
            differential (symbols) atomic pair distribution 
            functions for In$_{0.5}$Ga$_{0.5}$As.  The In-differential PDF 
            is significantly
	    broader than the total-PDF because $Q_{max}$ (as shown in
Fig.~\protect\ref{1}), and therefore
	    the real-space resolution of the measured PDF, is much lower.}    
\label{2}
\end{figure}

The high resolution total-PDF measurements and data 
analysis has been reported elsewhere \cite{[18],[jeong]}. 
Here, the experiment procedures and data analyses 
employed are reported in some more detail. The 
experiments were carried out at the A2 24 pole wiggler 
beam line at Cornell High Energy Synchrotron Source 
(CHESS). All measurements were done in a symmetrical 
transmission geometry at 10 K. The polychromatic 
incident beam was dispersed using a double crystal 
Si(111) monochromator and x-rays of energy 60 keV 
($\lambda=0.206$~\AA ) were employed. An intrinsic Ge detector 
coupled to a multi-channel (MCA) analyzer was used to 
detect the scattered radiation. By setting proper energy 
windows we were able to extract the coherent 
component of the scattered x-ray intensities during data
collection. The diffraction data were collected in 
scanning at constant $\Delta Q$ steps of 0.02~\AA\sp{-1}. Several runs 
were conducted and the resulting spectra averaged to 
improve the statistical accuracy  and reduce any 
systematic error due to instability in the experimental 
set-up. The diffraction data were smoothed using the 
Savitzky, Golay procedure \cite{[19]}. The procedure was 
tuned in such a way that each data point gained or lost only 
one Poisson counting standard deviation in the 
smoothing process.  The data were normalized for flux, 
and corrected for background scattering and 
experimental effects such as detector deadtime and 
absorption.  The part of Compton scattering at low 
values of $Q$ not eliminated by the preset energy window 
was removed analytically applying a procedure 
suggested by Ruland \cite{[20]}.  The resultant intensities were 
divided by the square of the average atomic form factor 
for the sample to obtain the total structure factor $S(Q)$,
\begin{equation}
 S(Q)=1+{\left[ I^{coh}(Q) - \Sigma c_if_i^2(Q)\right] \over
\left[\Sigma c_i f_i(Q)\right]^2}
\end{equation}
where I\sp{coh} is the coherent part of the total diffraction 
spectrum;  $c_i$ and $f_i(Q)$ are the atomic concentration and 
 scattering factor of the atomic species of type $i$ ($i= $
In,Ga,As), respectively \cite{[14],[16]}.  All data processing 
procedures were done with the help of the program RAD 
\cite{[21]}.  The reduced structure factor $F(Q)=Q[S(Q)-1]$ is shown in Fig.~\ref{1}.
As can be seen in the figure 
the F(Q) data are terminated at Q\sb{max} = 45~\AA\sp{-1} beyond 
which, despite the high intensity synchrotron source 
employed and extra experimental data averaging applied, 
the signal to noise ratio became unfavorable. It should be 
noted, however, that this is a very high value of the 
wavevector for an x-ray diffraction measurement; for 
comparison Q$_{max}$ achieved with a conventional source 
such a Cu anode tube is less than 8\ia . The 
corresponding reduced atomic distribution function, 
$G(r)$, obtained through a Fourier transform 
\begin{eqnarray}    G(r) =&&  4\pi r[\rho (r) - \rho_o]\nonumber \\
            =&& (2/\pi)\int^{Q_{max}}_{Q=0} Q[S(Q)-1]\sin Qr\>dQ,   
\end{eqnarray}
is shown in Fig.~\ref{2}.
Here $\rho (r)$ and $\rho_o$ are the {\it local} and 
{\it average} atomic number densities, respectively, and $r$ the 
radial distance. It should be noted that no modification 
function, i.e. additional damping of the S(Q) data at high 
values of $Q$, was carried out prior to Fourier transformation.
This ensures that the data have the highest resolution possible
but it results in some spurious ringing in G(r).  If Q$_{max}$
is high enough the ringing is small (on the level of the random
noise), and in any case it is properly modeled by convoluting 
G(r) with a Sinc function~\cite{[billinge]} which we do in all our 
modeling.

\subsection{Anomalous x-ray diffraction experiments at the In 
Edge}

It is well known that a single diffraction experiment on
an $n$-component system yields a total structure factor 
which is a weighted average of $n(n+1)/2$ distinct partial 
structure factors, i.e.
\begin{equation}
       S(Q) = \sum^n_{i,j} w_{ij}(Q)S_{ij}(Q),
\end{equation}
where $w_{ij}(Q)$ is a weighting factor and $S_{ij}(Q)$ the partial 
structure factor for the atomic pair ($i,j$), respectively 
\cite{[14]}. The corresponding total PDF, too, is a weighted 
average of  $n(n+1)/2$ partial pair correlation 
functions. For a multi component system like 
In\sb{0.5}Ga\sb{0.5}As it is therefore difficult to extract information 
about a particular atomic pair from a single experiment. 
The combination of a few conventional experiments (let 
say a combination of neutron, x-ray and electron 
diffraction experiments) or the application of anomalous x-ray 
scattering allows the determination of chemical-specific 
atomic pair distributions. 

\begin{figure}[tb]
    \centering
  \includegraphics[angle=0,width=3.0in]{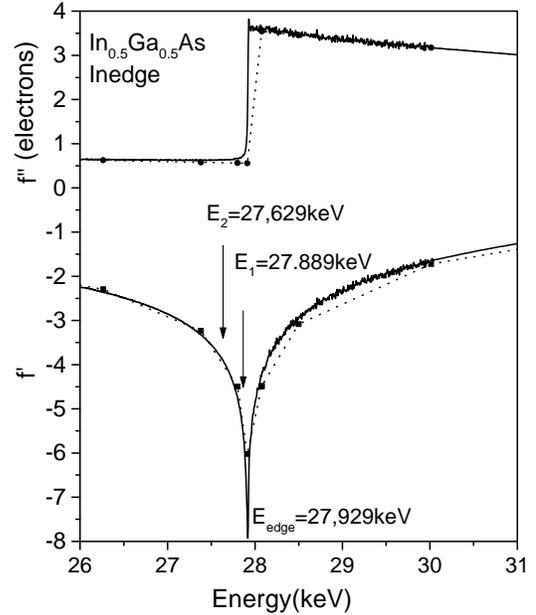}
\caption{            Energy dependence of the real, $f^\prime$, and 
           imaginary, $f^{\prime\prime}$,  anomalous scattering terms  of atomic 
           x-ray scattering factor of  In.  (Theoretical data - 
           symbols. The dotted line through the symbols is a 
           guide to the eye. Experimental data - full line). The 
           two energies below the In edge used in the present 
           anomalous scattering experiment are marked by 
           arrows. 
      }
\label{3}

\end{figure}

\begin{figure}[tb]    
\centering
  \includegraphics[angle=0,width=3.0in]{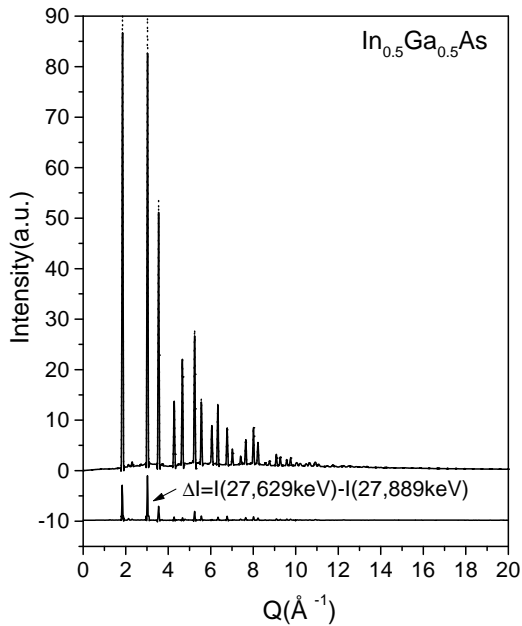}
\caption{
             Coherently scattered intensities of
            an In\sb{0.5}Ga\sb{0.5}As sample measured at two different 
            energies just below  the In edge (dots and solid line, 
            respectively). The difference, $\Delta I$, (offset for clarity) 
            between the two data sets is given in the lower part. 
}
\label{4}
\end{figure}

We briefly describe the use of anomalous scattering to obtain
chemical specific PDFs~\cite{[price]}.  If the incident x-ray photon energy is 
close to the energy of an absorption edge of a specific 
atom in the material, the atomic scattering factor should 
be considered a complex quantity dependent on both 
wavevector $Q$ and energy $E$
\begin{equation}
       f(Q,E) = f_o(Q) + f^\prime(E) + if^{\prime\prime}(E),                             
\end{equation}
where $f_o(Q)$ is the usual atomic scattering factor and $f^{\prime}$ 
and $f^{\prime\prime}$ are the anomalous scattering terms depending on 
the x-ray photon energy $E$.  The imaginary term, $f^{\prime\prime}$, is 
directly related to the photoelectric absorption 
coefficient and it is small and slowly varying for $E$ 
below the edge, rises sharply at the edge, and then 
gradually falls off. $f^{\prime}$ has a sharp negative peak at the 
edge with a width which is typically 40-80 eV at half 
maximum and is small elsewhere \cite{[22]}. This behavior can be clearly 
seen in Fig.~\ref{3}. 
The anomalous scattering 
technique takes advantage of the fact that $f^{\prime}$ and $f^{\prime\prime}$ for a 
particular atomic species change rapidly only within
$\sim 100$~eV of the respective absorption edge and that the 
characteristic absorption edges for different atomic 
species are separated by several hundreds of eV.  Then a 
difference of two sets of diffraction data taken at two 
slightly different energies below an absorption edge of a 
particular element will contain only a contribution of 
atomic pairs involving  that element. Accordingly, one 
can define a differential structure factor, DSF, as 
follows:
\begin{equation}
DSF= { I(E_1)-I(E_2)-\left[\Sigma c_i f^2(E_1) - \Sigma c_if^2(E_2)\right] \over
            \left[\Sigma c_i f(E_1)\right]^2 - \left[\Sigma c_if(E_2)\right]^2 }, 
\end{equation}    
where $E_1$ and $E_2$ are the two photon energies used 
\cite{[23]}. The differential PDF, which gives information 
about the atomic distribution around the anomalous 
scattering atoms, is calculated analogous to Eq. (3) with 
S(Q) replaced by the DSF. Several experiments have 
already demonstrated the usefulness of anomalous 
scattering techniques in studying the local atomic 
ordering in both disordered and crystalline materials \cite{[24],[25],[26]}.
  A precise knowledge of the anomalous scattering 
terms is, however, a prerequisite for the interpretation of 
anomalous scattering experiments. Unfortunately, 
theoretical models are not capable of providing precise 
enough values for $f^{\prime}$ and $f^{\prime\prime}$ in the vicinity of absorption 
edges. That is why anomalous scattering experiments 
usually involve a complimentary determination of  $f^{\prime}$ and 
$f^{\prime\prime}$.  It is most frequently done by measuring the energy 
dependence of $f^{\prime\prime}$ in the vicinity of the absorption edge 
and a subsequent determination of $f^{\prime}$ through the so-
called dispersion relation \cite{[27]}:
\begin{equation}
f^{\prime}(E_o ) = (2/\pi)\int {f^{\prime\prime}(E)\over[E_0^2 - E^2]}\,dE
\end{equation}                 
The same strategy was adopted in the present anomalous 
diffraction experiments. These were carried out at the In 
edge which is the highest energy ($\sim 27.929$~KeV) edge 
accessible in In-Ga-As system. The experiments were carried out at 
X7A beam line at the National Synchrotron Light Source 
at Brookhaven National Laboratory. Two energies, one 
just below (27.889 KeV) , and the other few hundred eV 
(27.629 KeV) below the In edge were used. The raw data are 
shown in Fig.~\ref{4}.
 A Si channel-cut monochromator 
was used to disperse the white beam. The 
monochromator was calibrated by measuring  the 
absorption edge of  indium of high purity. The scattered 
x-rays were detected by a Ge solid state detector 
coupled to a multi-channel analyzer. Few energy 
windows, covering several neighboring channels, were 
set up to obtain counts integrated over specific energy 
ranges during the data-collection. These energy windows 
covered the coherent intensity; coherent, 
incoherent and In $K\beta$ fluorescence intensities all 
together;  In $K\alpha$ fluorescence; As $K\alpha$ fluorescence, 
and a window covering the entire energy range which integrates
the total scattered intensity in the detector. Integrated counts of these 
ranges were collected several times and then averaged to 
improve the statistical accuracy. The data were corrected 
for detector dead time, Compton and background 
scattering, attenuation in the sample and residual In $K\beta$ 
fluorescence which is not possible to be well separated 
from the coherent component of scattered intensities.  In 
$K\beta$ was determined by monitoring the In $K\alpha$ signal and 
multiplying it by the ratio of In $K\beta$ to In $K\alpha$ output, 
which was experimentally determined by a 
complimentary experiment carried out well above the In 
edge ($\sim 29$ keV).
  
By taking the difference between the two 
data sets, as shown in Fig.~\ref{4}, 
all terms that do not 
involve In were eliminated, because only In scattering 
factor changed appreciably in the energy region explored 
while the scattering factors of Ga and As remained 
virtually the same, and the In DSF was obtained. 

The unknown anomalous scattering terms of In, 
involved in Eg. 5, were determined in the following way: 
In fluorescence yield was detected by scanning over a 
wide range across the In edge. The curve was matched to 
the theoretical estimates of Chantler \cite{[22]} for $f^{\prime\prime}$ and so 
the fluorescent yield converted to $f^{\prime\prime}$ data. $f^{\prime}$ was 
calculated from these $f^{\prime\prime}$ data via the dispersion relation 
as given in Eq. 6. The anomalous scattering terms of In, 
resulted from the present experiments, are given in Fig.~\ref{3}.  
As one can see in the figure,  in the vicinity of In 
edge $f^{\prime}$ and $f^{\prime\prime}$ change sharper than theory predicts. We 
determined the following values for $f^{\prime}$ and $f^{\prime\prime}$ for the two 
energies employed: $f^{\prime} = -3.89$ and $f^{\prime\prime}=0.637$ at $E=27.629$ 
keV;  $f^{\prime}=-6.148$ and $f^{\prime\prime}=0.826$ at E=27,889 keV. The use 
of the experimentally determined but not the 
theoretically predicted  values of $f^{\prime}$ and $f^{\prime\prime}$ turned out to 
be rather important in obtaining differential structure 
data of good quality. 
The  In-DSF and differential PDF for 
In\sb{0.5}Ga\sb{0.5}As alloy are shown in Figs.~\ref{1} and \ref{2}, 
respectively. The In-DSF appears broader primarily because $Q_{max}$, and 
therefore the resolution of the measurement, is much lower than is the
case for the total-PDF measurement (as is obvious in Fig.~\ref{1}).
There is an additional broadening of this peak because the In-DSP data
were collected a room temperature instead of 10K, but this is expected to
be small.

\section{Results}

As can be seen in Fig.~\ref{1} significant Bragg scattering (well 
defined peaks) is present up to approximately 15~\ia\ in the 
In difference and total structure factors of In\sb{0.5}Ga\sb{0.5}As 
alloy.  At higher wavevectors only an oscillating diffuse 
scattering is evident. This implies that although the 
sample still has a periodic structure it contains 
considerable local displacive disorder. The disorder is 
due to the mismatch of Ga-As and In-As bond lengths 
clearly seen as a split first peak in the total PDF of Fig.~\ref{2} 
\cite{[18]}. Also shown in Fig.\ref{2} is the In difference PDF 
which has a single first peak well lining up with the 
higher-r component of the first peak in the total PDF.  
Since the In differential PDF contains only atomic pairs 
involving In its first peak can be unambiguously 
attributed to In-As atomic pairs. This allows us to 
identify the two components of the first peak in the total 
PDF as being due to Ga-As and In-As atomic pairs, 
respectively.  According to the present high resolution 
x-ray diffraction experiments Ga-As and In-As bond 
lengths in the In\sb{0.5}Ga\sb{0.5}As alloy are 2.455(5) \AA\ and 
2.595(5) \AA\ at 10 K, respectively. In the present 
anomalous diffraction experiments In-As bond length is 
2.615(5) \AA\ at room temperature. The observed 
elongation of the In-As bond with temperature is due to the 
usual thermal expansion observed in materials. We note 
that the present PDF-based results are in rather 
good agreement with the XAFS results of Mikkelson and 
Boyce \cite{[4]} for Ga-As and In-As bond lengths in 
In\sb{0.5}Ga\sb{0.5}As. 

An inspection of the experimental PDFs in 
Fig.~\ref{2} (see also Figs.~\ref{9} and~\ref{7})
shows that the nearest atomic neighbor peak is the 
only one which is relatively sharp.  Starting from the 
second-neighbor peak onwards all atomic-pair 
distributions (PDF peaks) show significant broadening. 
The observation shows that the bond-length mismatch  
gives rise to a considerable deformation of the 
underlying zinc-blende lattice of In\sb{0.5}Ga\sb{0.5}As alloy.  To 
quantify this deformation we explored a few structure 
models as follows: 

\subsection{Supercell model based on the Kirkwood 
potential}

It was previously shown that a 512 atom supercell for the 
alloy structure, based on the zinc-blende unit cell but 
with In and Ga randomly arranged on the metal 
sublattice and atomic positions relaxed using the 
Kirkwood potential \cite{[32]}, explains well the 
high-resolution total-PDFs \cite{[18]}.  In addition to this high 
spatial resolution PDF we now have a PDF which is 
chemically resolved.  We are interested to know whether 
this supercell model is still sufficient for describing these 
new data.  The model has been described in detail 
elsewhere \cite{[chung]}.

\begin{figure}[tb]
\centering
  \includegraphics[angle=0,width=3.0in]{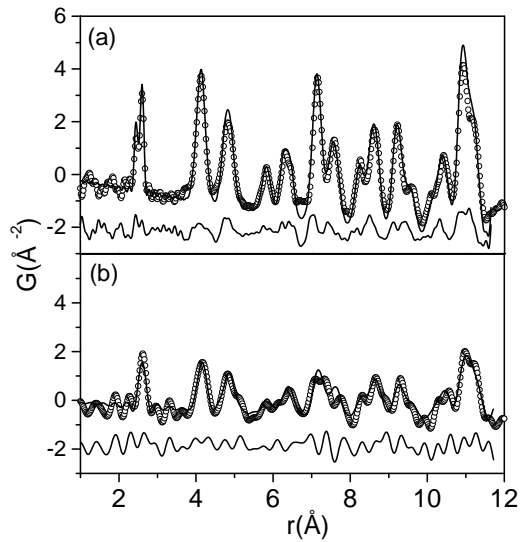}
\caption{
  Experimental (open circles) and model 
    (solid line) PDFs for In\sb{0.5}Ga\sb{0.5}As alloy. (a)  total PDF; 
    (b)  In differential PDF. Model PDFs are based on 
    Kirkwood-type potential minimization. The residual 
    difference between the model and experimental data is 
    given in the lower part. 
}
\label{9}
\end{figure}

In the present modeling we used the same force 
constants  $\alpha_{ij}$  and $\beta_{ij}$ that were selected to fit the end 
members GaAs and InAs \cite{[18],[chung]}. Using the relaxed 
atomic configuration a dynamical matrix has been 
constructed and the eigenvalues and eigenvectors found 
numerically. From this the Debye-Waller factors for all 
the individual atoms in the supercell have been 
determined. The PDF of the model was then calculated 
using a Gaussian broadening of the atomic-pair 
correlations to account for the purely thermal and zero-point 
motion.  The width of the Gaussians was determined 
from the theoretical Debye-Waller factors~\cite{[chung]}.  In addition, 
the calculated PDF was convoluted with a Sinc function to account 
for the truncation of the data at $Q_{max}$.~\cite{[billinge]}.  A 
comparison between the model and experimental results 
is shown in Fig.~\ref{9}. 
The agreement with both the high
spatial resolution data (Fig.~\ref{9}(a)) and with the 
differential PDF (Fig.~\ref{9}(b)) is clearly very good. It has already 
been demonstrated \cite{[18]} that the Kirkwood-based model well 
reproduces the displacements of As and metal atoms in 
In-Ga-As alloys extracted from a model independent 
analysis the PDF peak widths. Thus the present and 
previous results suggest that the Kirkwood-based model is a 
good starting point for any further calculations requiring 
good knowledge of the local structure of  the In\sb{0.5}Ga\sb{0.5}As 
alloy.   

\subsection{Refinement of chemically resolved differential 
and Spatially resolved Total PDF}

In this paper we present, for the  first time, both high-resolution total-
and chemically resolved In-partial PDFs for In$_{0.5}$Ga$_{0.5}$As.  
In the previous Section
we showed that the data are consistent with a supercell model
 based on the Kirkwood potential.  However, in addition we would like 
to extract structural information from the data without recourse
to potential-based models which can be used 
to compare with other experimental 
results~\cite{[4],[9],haga;apl85,fukui;jjap84,glas;pmb90,landa;ssc93,glas;pma87} 
and theoretical 
predictions~\cite{[5],[8],[9],silve;epl95}.

To do this we have
constructed the simplest possible model that was still
consistent with the data, and we have refined it using the PDF 
profile-fitting program PDFFIT~\cite{[28]}.  We have fit to 
both the high-spatial resolution total-PDF {\it and} the chemically 
resolved differential PDF data at the same time
which resulted in equivalent  atomic
displacement parameters being refined.

The model is based on the 8-atom cubic unit cell of the 
zinc-blende structure.
The split nearest-neighbor peak in the total PDF, 
and the shifted nearest-neighbor peak in the In-differential PDF,
both require that definite static displacements of fixed length
be incorporated in the model. In addition, the shifted position
of the  nearest neighbor peak in the In-differential PDF requires that a model be
constructed which has a definite chemical species on specific sites,
i.e., goes beyond the virtual crystal approximation.  In simple
8-atom cubic unit cell this {\it de facto}
leads to a chemically ordered model that is not observed in the
real alloy and which, furthermore, does not sample all of the 
possible chemical environments for As in the random 
alloy~\cite{[9],silve;epl95,schab;prb91}.  Nonetheless, this is the minimal
model which can be successfully refined to the experimental data to 
extract information about local atomic displacement amplitudes.

\begin{figure}[tb]
\centering
  \includegraphics[angle=0,width=3.0in]{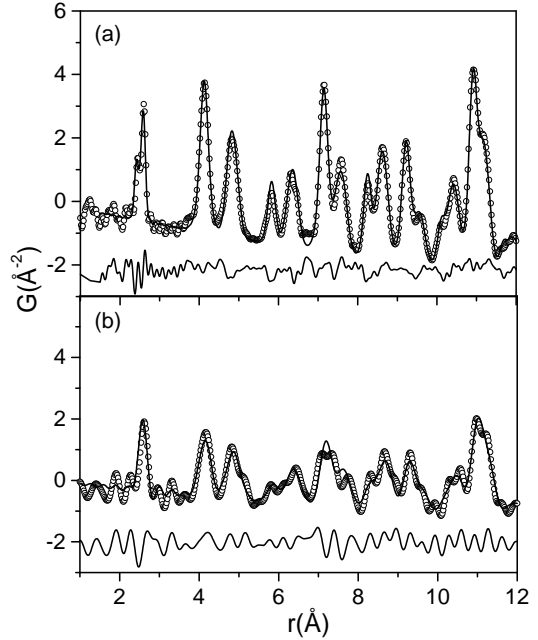}
\caption{Experimental (open circles) and model (solid 
 line) PDFs for In\sb{0.5}Ga\sb{0.5}As alloy. (a) – total PDF; (b) – 
 In differential PDF. Model PDFs are based on 
 a distorted zinc-blende-type structure. The residual 
 difference between the model and experimental data is 
 given in  the lower part. 
}
\label{7}
\end{figure}

In this model the four metal sites are populated with two In and
two Ga ions. Static displacements of As and metal ions were then allowed.
The model was constrained so that all four As sites had the same displacement
amplitude.  The metal sites were likewise constrained to be displaced
by the same amount as each other, but the metal site displacement
was independent of the As sublattice displacement.
The directions of the displacements were also constrained to be along
 $\langle 111\rangle$ type directions. The
choice of which of the 8 possible  $\langle 111\rangle$ 
directions  was determined by the
chemical environment.   
A model with $\langle 100\rangle$ type 
displacements was less successful at reconciling the sharp first peak
and broad later peaks in the PDF.  This is discussed in more detail later.
We call these the ``discrete'' displacements.
In addition to the discrete atomic 
displacements, atomic-displacement-parameters
(thermal factors) and lattice parameters
were refined.  The thermal factors contain both {\it static} and {\it dynamic}
disorder.  These atom displacement distributions we refer to as
``continuous'' to differentiate from the discrete displacements
described above.
Finally, the nearest neighbor peak was sharpened with respect
 to the rest of the PDF using a sharpening factor.  This accounts for 
the highly correlated nature of the displacements of near-neighbor 
atoms.~\cite{jeong;jpca99}
The resulting fit to
the data is shown in Fig.~\ref{7}.

The values we refine are as follows: the discrete displacements on
the As and Metal sublattices are 0.133(5)~\AA\ and 0.033(8)~\AA\ respectively.
These values are independent of temperature.  The continuous-displacement
amplitudes are 
$\langle u^2_{As}\rangle = 0.00814(12)$~\AA$^2$\  
and $\langle u^2_{M}\rangle = 0.00373$~\AA$^2$\  for the 
As and metal sublattices at 
$T=10$~K and $\langle u^2_{As}\rangle = 0.0135(15)$~\AA$^2$\  
and $\langle u^2_{M}\rangle = 0.010(2)$~\AA$^2$\  , 
respectively, at room 
temperature.  These
compare with literature values of 
$\langle u^2_{As}\rangle = 0.0015(8)$~\AA$^2$\  and 
$\langle u^2_{M}\rangle = 0.0017(9)$~\AA$^2$\  for the end-member 
compounds at 10~K~\cite{[18]}, and
$\langle u^2_{As}\rangle = 0.00716(5)$~\AA$^2$\  and 
$\langle u^2_{M}\rangle = 0.009$~\AA$^2$\  at room 
temperature~\cite{stahn;acb98} 
for As and 
the metal site, respectively.

\begin{figure}[tb]
\centering
  \includegraphics[angle=0,width=3.0in]{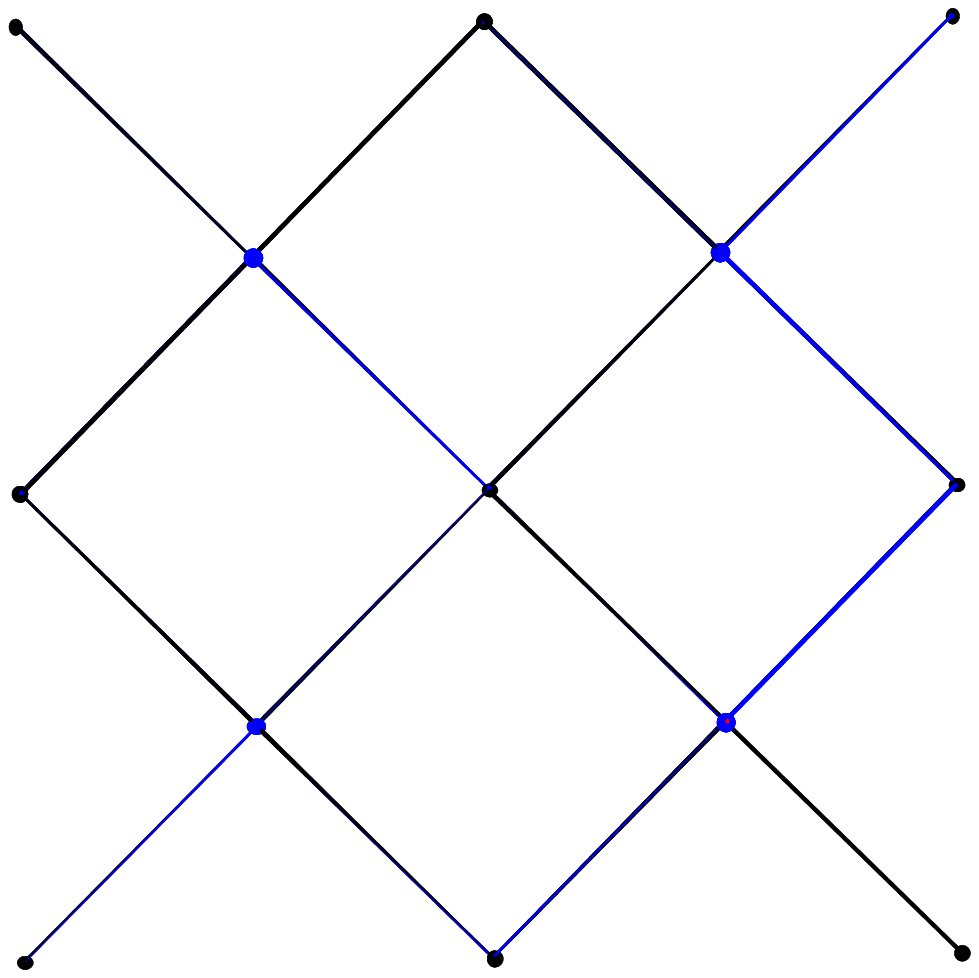}
  \includegraphics[angle=0,width=3.0in]{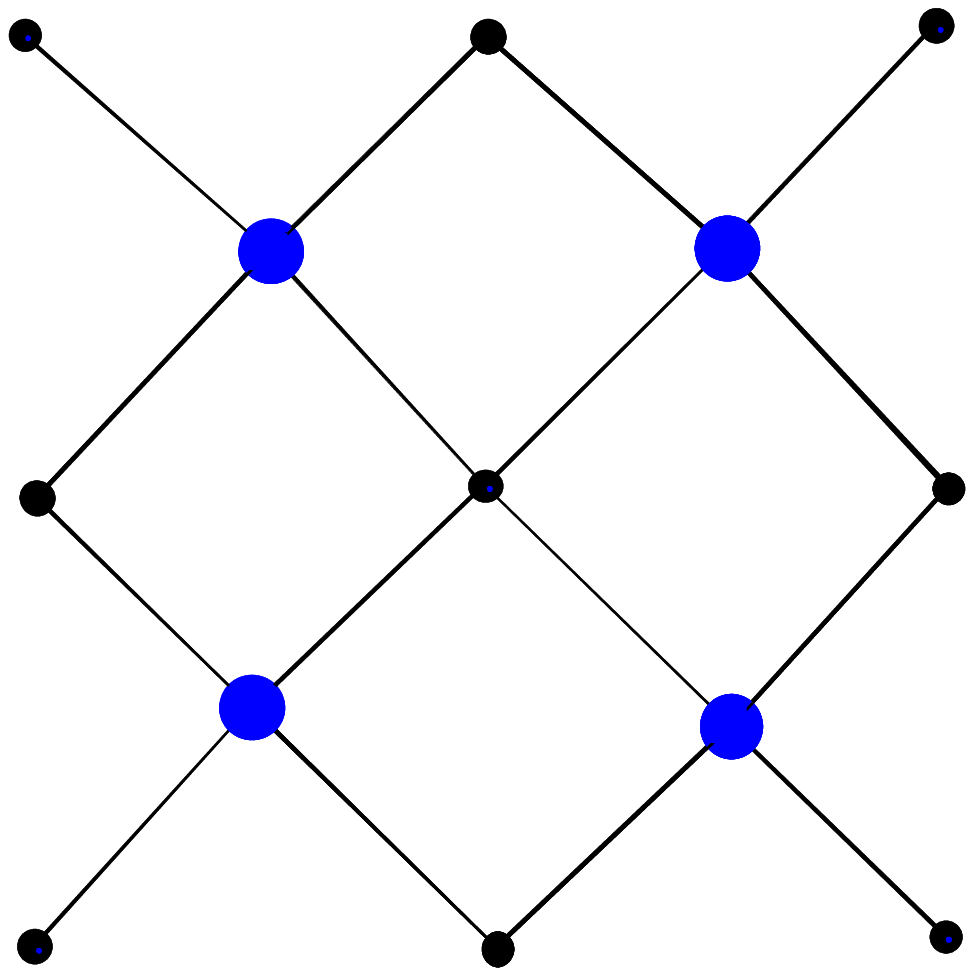}
\caption{Schematic representation of the discrete distortions to the zinc-blend
structure obtained from the fitting.
Fragments from the ideal structure (top panel) can be compared with the
distorted (lower panel) structure.  The view is a projection along the
[010] direction.  The large circles are the As sites
and the small circles the metal sites.  The relative sizes of the circles 
reflect schematically the relative sizes of the continuous-displacements on 
each site in the undoped material (top panel) and the alloy (bottom);
however, their size has been exaggerated.
}
\label{fig;models}
\end{figure}

The discrete displacements obtained from the fits are 
illustrated schematically, 
by a projection of a fragment of the structure
down the [010] direction, in Fig.~\ref{fig;models}.
This shows that both the discrete and continuous displacements (indicated
schematically by the size of the circles representing the atoms) are larger
on the As than the metal sublattice.  The size of the circles
representing the continuous-displacements have been exaggerated.

\section{Discussion}

Existing experimental results that characterize the structure of
semiconductor alloys beyond the 
average structure include XAFS~\cite{[4],[9],fukui;jjap84}, 
ion-channeling~\cite{haga;apl85}, x-ray diffuse 
scattering~\cite{glas;pmb90}, and Raman scattering~\cite{landa;ssc93}.  
The XAFS results clearly show that
short and long near-neighbor bonds exist which are from Ga-As and In-As
neighbors respectively.  The bond-length {\it distribution} is not
recovered with great accuracy and there is limited information available
on higher neighbor pairs; however, the indication is that the atom-pair
separations return quickly to the virtual crystal values with increasing
pair-separation, $r$.  This implies significant distortions to the
crystal structure.
Indeed, a correct analysis of the phonons in Raman  spectra from 
Ga$_{1-x}$In$_x$As required significant structural 
distortions~\cite{landa;ssc93}. 
Our current and earlier~\cite{[18]} 
PDF results bear out all these observations.

The discrete displacements refined in our $\langle 111\rangle$
displaced model are primarily determined by the splitting, and
displacement, of the first peak in the total and In-differential PDF's,
respectively: this sharp feature in the PDF is very sensitive to the
amplitude of the discrete displacement.  The bond-length difference,
$\Delta r$,
between the end-members is 0.173~\AA\ and is $\sim 0.14$~\AA\ in
the alloy~\cite{[4],[18]}.  If we add the discrete displacements on the arsenic
and metal sites we get $\Delta r=0.16(2)$~\AA .  The bond-length
difference can be obtained directly by fitting Gaussians to the
first PDF peak in a model independent way~\cite{[18]}.  What this modeling
shows is that within the structure, most of the relaxation of local
bonds occurs by arsenic moving off its site, but displacements
of the metal atoms are also important.

Ion channeling results give
a precise determination of the mean-square displacement amplitude, 
$\langle u^2\rangle$, 
(including static and dynamic components)
perpendicular to the channeling direction. The results of 
Haga~{\it et al.}~\cite{haga;apl85} give a $\langle u^2\rangle$
for Ga$_{0.47}$In$_{0.53}$As of 0.017~\AA$^2$\ at room temperature.
Based on the theoretical thermal amplitude of the end-member compounds being
$\langle u^2\rangle=0.0121$~\AA$^2$, (and in good agreement with 
the value measured using the PDF~\cite{[jeong]}) they determined that the
mean-square static displacements perpendicular to [100] were of magnitude
$\langle u^2\rangle=0.005$~\AA$^2$.  They saw a similar value in
directions perpendicular to $\langle 110\rangle$.  This corresponds to
a static root-mean-square displacement amplitude of 0.07~\AA .  
This is much smaller than the discrete displacements of 0.133~\AA\ that
we observe.  However, if we find the average displacement by 
adding the discrete displacements on the 
As and metal sublattices in quadrature and dividing by 2  we get
0.068~\AA$^2$ in good agreement with the ion channeling.
Their work
did not report which sublattice contributed most of the disorder;
however, there is a suggestion from electron 
diffraction~\cite{glas;pma87}, in agreement
with theory~\cite{[5],[8],[9],silve;epl95}, 
that the As sublattice is more disordered as we show directly
from our measurement.


Finally, we note that the actual displacement pattern on the arsenic
site is expected to have $\langle 111 \rangle$ type displacements
(as in our model) but also significant $\langle 100 \rangle$
type displacements~\cite{[9],[jeong],silve;epl95}.  In fact, recent
calculations indicate that the $\langle 100 \rangle$ displacements
should be significantly more pronounced than $\langle 111 \rangle$
displacements, especially at room temperature~\cite{silve;epl95}.
We are undertaking a more sophisticated modeling approach to 
explore this prediction.  We tried a simple $\langle 100 \rangle$ 
displaced model, analogous to the one described here, but found it
to explain the data less successfully than the $\langle 111 \rangle$
model we described. We feel that this is a deficiency of the simple
single-displacement-direction modeling rather than signifying that
the displacement directions are actually $\langle 111 \rangle$ in 
the real alloy.  The likely reason is that larger displacement
amplitudes are required 
along $\langle 100 \rangle$ directions to satisfy the
bond-length difference seen in the first PDF peak (actually 
$\sqrt{2}$ times larger).  With these large discrete displacements
it is harder for the model to account for additional disorder in
the data using enlarged thermal factors.  A better fit in these
imperfect models is obtained with
smaller discrete displacements coupled with larger continuous-displacements.
However, an improved fit should result from a more sophisticated model
which includes both $\langle 100 \rangle$ and $\langle 111 \rangle$
type displacements.

\section{Conclusions}

From high real space resolution total and In differential 
PDFs of In\sb{0.5}Ga\sb{0.5}As alloy we conclude the following:   
In good agreement with earlier XAFS results~\cite{[4]}
the Ga-As and In-As bonds do not take some 
compositionally averaged length but remain close to 
their natural lengths in In\sb{0.5}Ga\sb{0.5}As alloy.  This bond-
length mismatch brings about a considerable local 
disorder seen as a significant broadening of the next-
nearest atomic-pair distributions.  
The positions and widths of the low and high-$r$ peaks in
both the total- and indium differential-PDFs are very well
reproduced using a relaxed supercell model based on the
Kirkwood potential with parameters taken from fits to the
end-members of the alloy series.  This suggests that this is
a reasonable approach for generating the local structure of these
alloys.

A co-refinement of both the high-resolution total PDF {\it and}
the chemically resolved indium differential PDF using a
simplified structural model was carried out. The arsenic sublattice
contains most of the disorder in the structure as evidenced from
both discrete atomic displacements in the model and enlarged
thermal parameters.  However, small, but significant, displacements
are evident on the metal sites and these have been quantified.

\acknowledgements

We would like to acknowledge M. F. Thorpe and J. Chung
for discussions and for making the results of their 
supercell calculations available for comparison with the
experimental data.  We would like to acknowledge S. Kycia
and A. Perez for help in collecting the CHESS data.
We are very grateful to T. Egami for making x-ray beamtime
available to us and to D. E. Cox for help with the NSLS
experiments.
This work was supported by DOE grant DE FG02 
97ER45651. S.J.L.B. also acknowledges support from 
the Alfred P. Sloan Foundation as a Sloan Fellow.  High-
energy x-ray diffraction experiments were carried out at 
Cornell High Energy Synchrotron Source (CHESS). 
CHESS is operated by NSF through grant DMR97-
13424. Anomalous scattering experiments were carried 
out at The National Synchrotron Light Source, 
Brookhaven National Laboratory, which is funded under 
contract DE-AC02-98CH10886.

\end{document}